\documentstyle[prl,aps,multicol,epsf]{revtex}
\begin{document}
\draft

\title{Spin Polarization of Two-Dimensional Electrons Determined from Shubnikov-de
Haas Oscillations as a Function of Angle.}
\author{S. A. VITKALOV and M. P. SARACHIK} 
\address{Physics Department, City College of the City University of New
York,
New York, NY 10031}
\author{T. M. KLAPWIJK}
\address{Department of Applied Physics, Delft University of Technology,
2628 CJ Delft, The Netherlands}
\date{\today}
\maketitle

\begin{abstract}
Recent experiments in the two dimensional electron systems in silicon MOSFETs have
shown that the in-plane magnetic field $H_{sat}$ required to saturate the conductivity
to its high-field value and the magnetic field $H_s$ needed to completely
align the spins of the electrons are comparable. By small-angle Shubnikov-de Haas
oscillation measurements that allow separate determinations of the spin-up and
spin-down subband populations, we show to an accuracy $5$\% that $H_{sat}=H_s$.
\end{abstract}
\vspace {6mm}
\begin{multicols}{2}

\noindent \underline{}
\vspace{0.2in}

Dilute two-dimensional electron (or hole) systems which display unexpected metallic
behavior are currently the subject of great interest \cite{review}.  These strongly
interacting systems also exhibit enormous magnetoresistances in response to a
magnetic field applied parallel to the electron plane
\cite{dolgopolov,simonianH,pudalovH,simmons,yoon}: with increasing field, the
resistivity rises sharply and saturates to a constant value for
$H>H_{sat}$; the saturation field $H_{sat}$ is of the order of several tesla and
varies with temperature and electron density.

Two experiments have recently shown
that the behavior of the resistance is related to the spin polarization of the
electrons.  On the assumption that the product $gm^*$ of the interaction-enhanced
g-factor $g$ and effective mass $m^*$ measured in small field does not change in a
strong in-plane magnetic field, Okamoto {\it et al.} \cite{okamoto} found that the
saturation field $H_{sat}$ is close to the field at which the 2D electron system in
silicon MOSFETs becomes fully spin polarized.  Similar conclusions were drawn by
Vitkalov {\it et al.} \cite{vitkalov}, who demonstrated that the frequency of small
angle SdH oscillations versus filling factor $\nu$ in high fields $H>H_{sat}$ is
double the frequency at low fields $H<H_{sat}$.  This signals a decrease of the
density of states by a factor of two and the complete depopulation of one of the spin
subbands when $H>H_{sat}$.  The equivalence between the field $H_{sat}$ at which the
resistance saturates and the field $H_s$ required to obtain complete alignment of the
electron spins was established in these experiments to an accuracy of $\approx 10$ to
$15$\%.  By using a modification of the small-angle Shubnikov-de Haas method of
measurement\cite{vitkalov} which allows a determination of the population of each
spin subband separately, we show in the present paper that $H_{sat}=H_s$ with accuracy
of$\approx 5$\% \cite{shayegan}.

The method is based on the following considerations.  The populations of the spin-up
and spin-down subbands are governed by the Zeeman energy which is determined by the
total magnetic field.  The  splitting of the Landau levels is controlled by the
normal component of the magnetic field $H_\perp$: $\hbar \omega_c=eH_\perp/mc$. 
By rotating the 2D electron plane relative to the direction of the magnetic field we
change the normal component of the magnetic field and, therefore, the Landau level
spitting in each band.  For a fixed total magnetic field $H$, the sizes of the Fermi
circles $k_F^{\uparrow, \downarrow}$, formed by the spin-up and spin-down electrons
are different and constant.  The frequency of the SdH oscillations with $1/H_\perp$
is proportional to the area of the Fermi circle and therefore the density of spin-up (or
spin-down) 2D carriers  Thus, the ratio of the frequencies of the SdH oscillations due
to the spin-up and spin-down Fermi circles yields the ratio of populations of the spin
subbands for a given value of in-plane magnetic field.  Analyzing the data obtained by
this method, we show that the saturation field $H_{sat}$ is the same as the
field required for complete spin polarization with an accuracy of $5$\%.

Measurements were taken on a silicon MOSFET; the mobility $\mu$ at $0.1$ K was 
$26,000\;$V/(cm$^2s)$.  Contact resistances were minimized by using a split-gate
geometry, which allows a higher electron density in the vicinity of the contacts
than in the 2D system under investigation.  Standard $AC$ four-probe techniques
were used to measure the resistance with $AC$ currents in the linear regime,
typically below 5 nA, at frequency 3Hz.  Data in high magnetic fields up to 20T
were obtained at the National Magnetic Field Laboratory in Tallahassee, Florida.  
The sample was mounted at the end of a low temperature probe on a rotating platform. 
The sample was rotated in constant magnetic field by a stepper motor.  The electron
density $n_s$ was fixed during the sample rotation and measurements were taken at a
temperature of about $100$ mK.  The longitudinal and the Hall voltages were detected
simultaneously.

For small angles $\phi$ between the electron plane and the magnetic field direction,
the Hall resistance is proportional to the normal component of the magnetic field
$H_{\perp}$: $R_{xy}=H_{\perp}/(n_se)$, with $H_{\perp} = H$sin$\phi$.  The Hall
coefficient does not depend on the degree of the spin polarization of the 2D
electrons in silicon MOSFETs.  This was demonstrated in recent experiments \cite{Hall}
with an accuracy $\approx 5$\% for electron densities below $n_s=2.75 \times 10^{11}$
cm$^{-2}$.  We note that the proportionality of $R_{xy}$ with $H_{\perp}$ is
sufficient to guarantee the correct $ratio$ between the two periods of the SdH
oscillations, regardless of whether the constant of proportionality varies with
in-plane magnetic field.

The angle between the magnetic field and the 2D plane was determined also from the known gearing number of the stepper motor.  Both methods give the same ratio of frequencies
with accuracy about $3$\%.  The filling factor was calculated, using
the relation
$\nu=n_s\Phi_0/H_{\perp}$.
\vbox{
\vspace{0.2in}
\hbox{
\hspace{-0.3in} 
\epsfxsize 3.3in \epsfbox{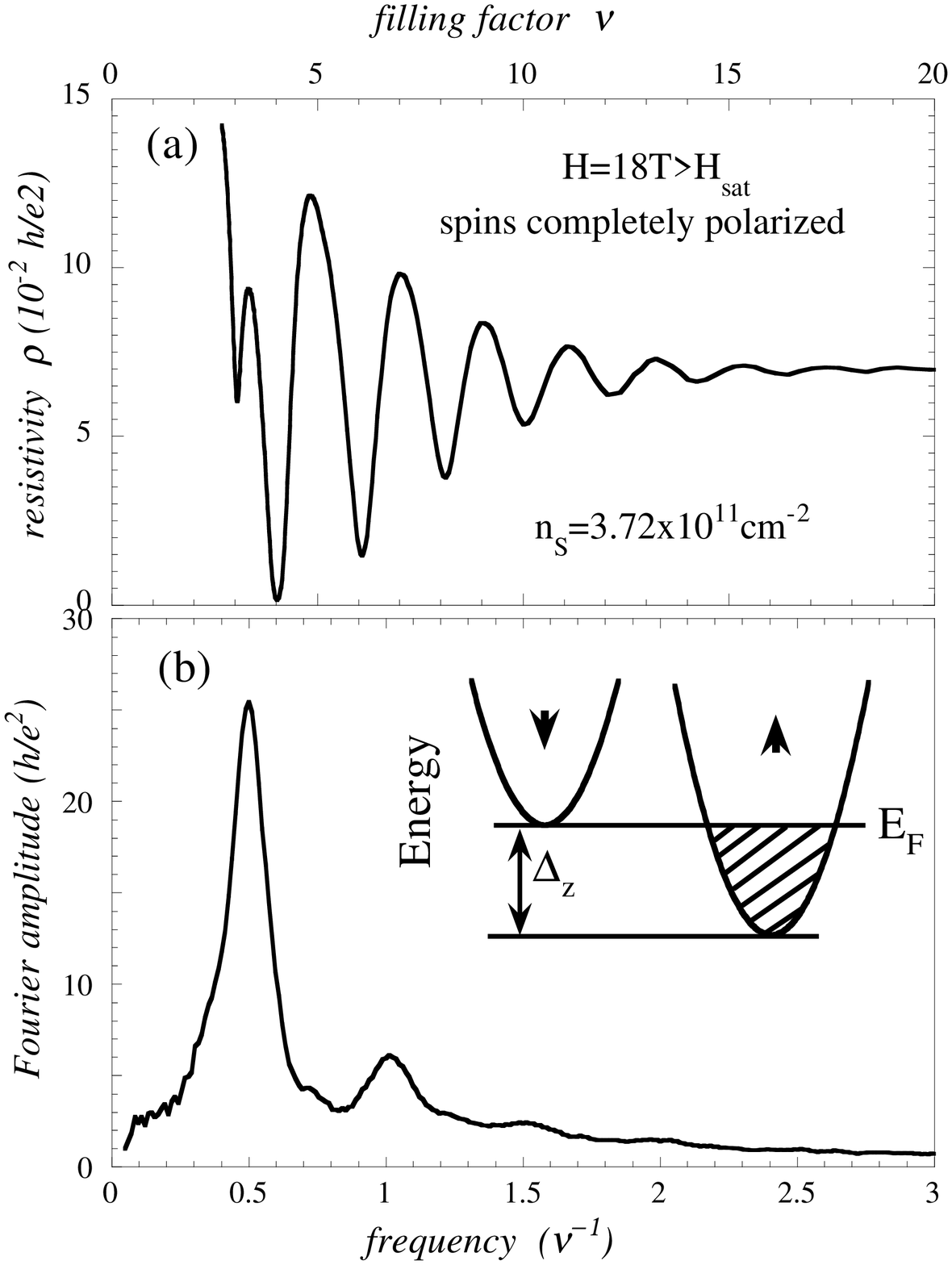} 
}
}
\refstepcounter{figure}
\parbox[b]{3.1in}{\baselineskip=12pt FIG.~\thefigure.
(a) Small-angle Shubnikov-de Haas oscillations of the two-dimensional system of
electrons in a silicon MOSFET at density $3.72 \times 10^{11}$ cm$^{-2}$ in
the presence of a magnetic field of $18$ T, which close to the saturation field $H_{sat}$ for this density.  
(b) Fourier transform of the data shown
in part (a); the inset is a schematic band diagram, corresponding to the complete spin polarization at $H=H_{s}$.
\vspace{0.10in}
}
\label{1}

\vbox{
\vspace{0.2in}
\hbox{
\hspace{-0.3in} 
\epsfxsize 3.3in \epsfbox{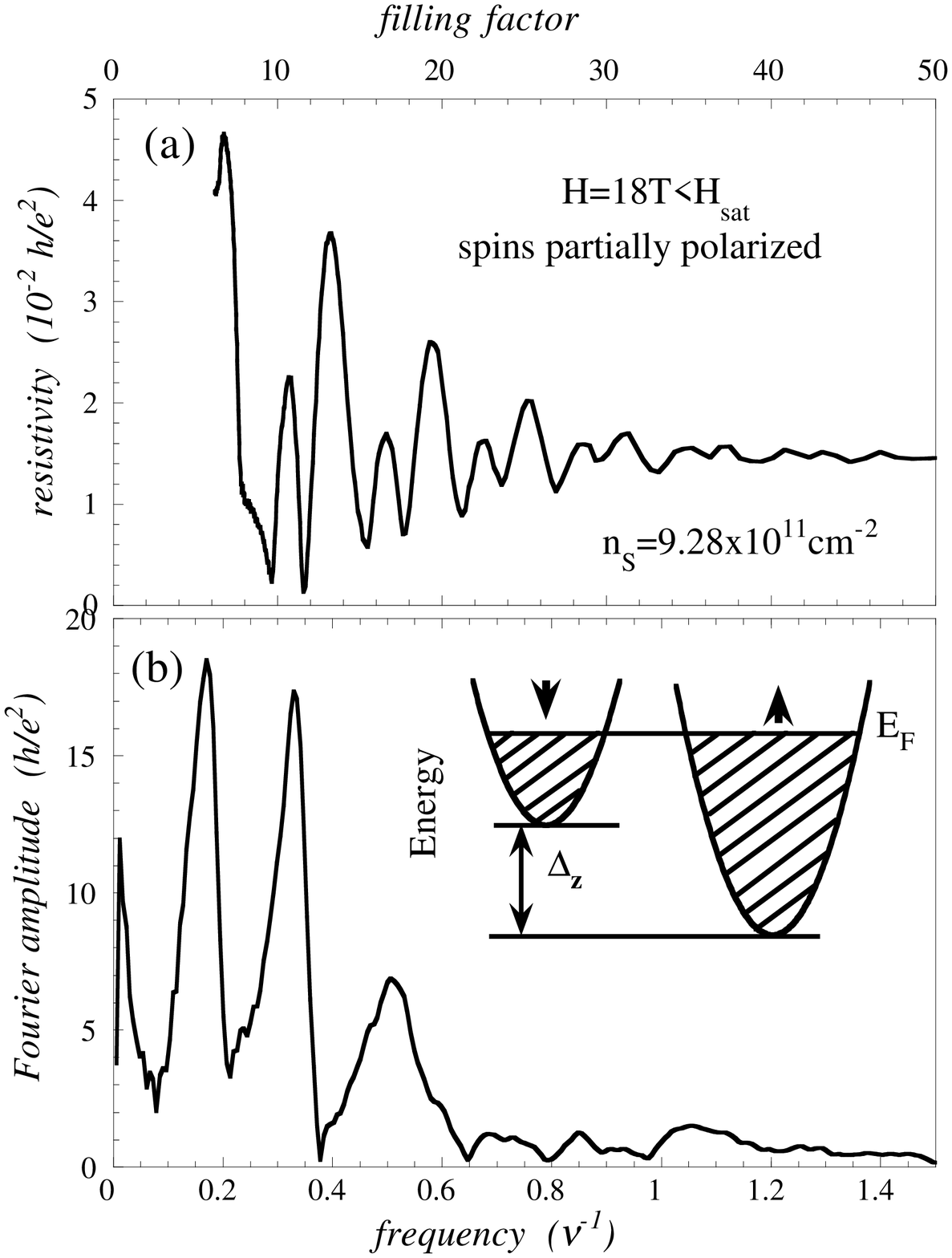} 
}
}
\refstepcounter{figure}
\parbox[b]{3.1in}{\baselineskip=12pt FIG.~\thefigure.
(a) Small-angle Shubnikov-de Haas oscillations of the two-dimensional system of
electrons in a silicon MOSFET at density $9.28 \times 10^{11}$ cm$^{-2}$ in
the presence of a magnetic field of $18$ T.  (b) Fourier transform of the data shown
in part (a); the inset is a schematic band diagram, corresponding to partial spin
 polarization of the 2D electrons at $H<H_{s}$.
\vspace{0.10in}
}
\label{2}

In a constant total magnetic field of $18$ tesla, the longitudinal resistivity is
shown in Fig. 1(a) plotted as a function of filling factor $\nu$ for electron
density $n_s=3.72 \times 10^{11}$ cm$^{-2}$.  Clean periodic Shubnikov de Haas
oscillations are observed with a period $\Delta \nu=2$.  Taking into account the
two-fold valley degeneracy of 2D electrons in silicon MOSFETs \cite{Ando}, the
period $\Delta \nu=2$ is found to correspond to quantum oscillations in a single
spin band.  In other words, at $H=18$ T and $n_s=3.72 \times 10^{11}$ cm$^{-2}$ the
electrons are spin polarized completely.  The sharp dip at $\nu=3$ corresponds to
valley spitting.  The Fourier spectrum of these oscillations is shown in Fig. 1(b). 
The main peak in the spectrum corresponds to the period $\Delta \nu=2$ which is
clearly dominant in Fig. 1(a).  The two smaller maxima at higher frequencies are
second and third harmonics.  Valley splitting, oscillations of the Fermi energy,
and various other effects may be responsible for these higher order peaks.  The
schematic band diagram in Fig. 1(b) shows the population of the spin up and
spin-down bands for full spin polarization and $H=H_{s}$, the condition that obtains
at the density shown.  A similar pattern of oscillations with  period $\Delta
\nu=2$ is observed for a lower electron density $n_s=2.75 \times 10^{11}$ cm$^{-2}$.

Shubnikov-de Haas oscillations are shown in Fig. 2(a) for the same magnetic field
$H=18T$ for a substantially higher density  $n_s=9.28 \times 10^{11}$ cm$^{-2}$. In
contrast with Fig. 1, the oscillations do not have a single period.  Instead they
demonstrate a beating pattern.  The main frequencies that create the beats are
easily identified in Fig. 2(b), which shows the Fourier components of the
oscillations.  The schematic band diagram in Fig. 2(b) shows the population of the
spin up and spin-down bands corresponding to the high-density case, where
$H<H_{sat}$ and the electons are partially polarized.  The two main peaks of Fig.
2(b) are at frequencies $0.167$ and $0.330$, corresponding to quantum oscillations in
the spin-down and spin-up subbands.  They are proportional to the Fermi energies (and
therefore the populations) of the spin-down and spin-up bands.  The Fourier spectrum
contains several additional peaks associated with obvious and strong nonlinearity of
the SdH spectrum.  A complete interpretation of the different peaks will require
detailed analysis.

The ratio of the frequencies of the oscillations in the two spin subbands is
proportional to the ratio of the densities of spin-up and spin-down electrons:
$f^\uparrow/f^\downarrow=n_s^\uparrow/n_s^\downarrow$.  The ratio
$f^\downarrow/f^\uparrow$ determined from our experiments is plotted in Fig. 3(a) for
different total electron densities $n_s$.  At high electron density  $n_s=9.28 \times
10^{11}$ cm$^{-2}$ the 2D electron system is partially spin polarized at $H=18T$ and
the ratio $f^\uparrow/f^\downarrow$ is $\approx 0.5$.  Decreasing the total
electron density $n_s$ reduces both the spin-up and the spin-down populations,
yielding a smaller $f^\uparrow/f^\downarrow$.  Finally, for
$n_s=3.72 \times 10^{11}$ cm$^{-2}$ and below, the spin-down subband is
depopulated completely at $18$ Tesla, and the 2D system is fully spin polarized.  The
ratio $f^\uparrow/f^\downarrow$ is equal to zero for $n_s < n_{sat}(18$ T$)=3.72
\times 10^{11}$ cm$^{-2}$.

The magnetic field $H_s$ required to achieve full spin polarization of the electron
system can be calculated from the following simple considerations.  The Fermi energy
of the spin-up and spin-down electrons measured relative to the bottom of each
subband is
$$
\epsilon^{\uparrow,\downarrow} = \epsilon_F^0 \pm \mu g H/2 = n/(2D) \pm \mu g H/2 
\eqno{(1)}
$$
where $\epsilon_F^0 = n/(2D)$ is the Fermi energy of each band when $H=0$, and the
density of states $D$ of the electrons in two dimensions is constant \cite{Ando}.  The
ratio
$$
\frac{f^\downarrow}{f^\uparrow} = \frac{n^\downarrow}{n^\uparrow} =
\frac{\epsilon^\downarrow D}{\epsilon^\uparrow D} = \frac{1-(g \mu H D)/n}{1+(g \mu H
D)/n}. 
\eqno{(2)}
$$
The condition for full spin polarization in a magnetic field $H_s$ is that
$n^\downarrow = 0$, thus $\epsilon^\downarrow = 0 = n/(2D) - \mu g H_s/2$, and $n/D
= \mu g H_s$.  Substitution \cite{assumption} into Eq. (2) yields:
$$
\frac{f^\downarrow}{f^\uparrow} = \frac{(1-H/H_s)}{(1+H/H_s),}
$$
an expression that is valid when $H < H_s$ ($H=18$ T in our experiments).  The field
$H_s$ can thus be calculated from the expression:
$$
H_s = \frac{(1+f^\downarrow/f^\uparrow)}{(1-f^\downarrow/f^\uparrow)}. H \eqno{(3)}
$$

\vbox{
\vspace{-0.2in}
\hbox{
\hspace{-0.3in} 
\epsfxsize 4in \epsfbox{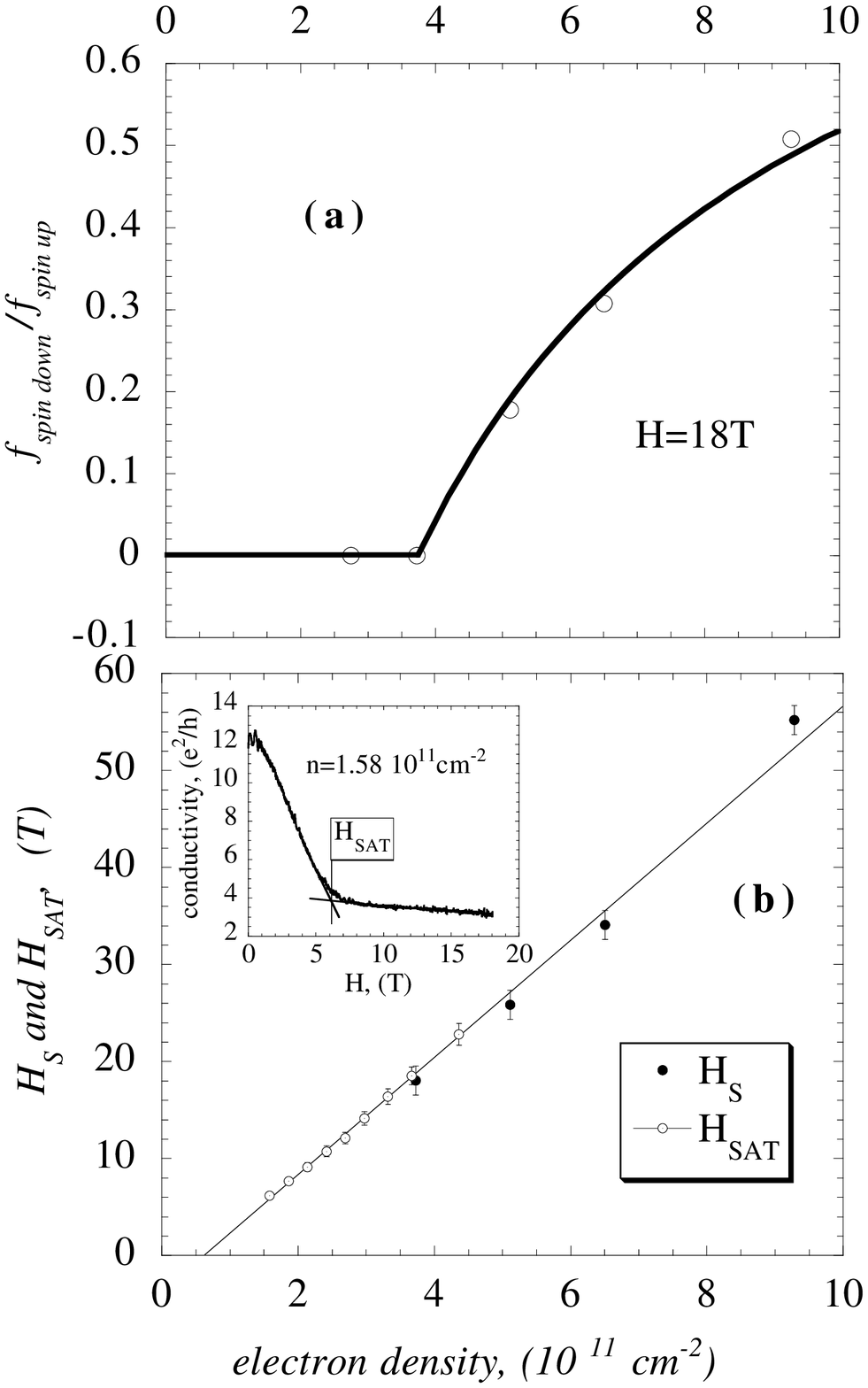} 
}
}
\refstepcounter{figure}
\parbox[b]{3.1in}{\baselineskip=12pt FIG.~\thefigure.
(a) The ratio $f^\downarrow/f^\uparrow$ versus electron density $n_s$ in a
magnetic field $H=18$ T applied at a small angle relative to the plane of a
silicon MOSFET; the line is drawn to guide the eye.  (b)  Open circles denote
the field $H_{sat}$ at which the conductivity saturates to its high-field value,
as shown in the inset.  The closed circles are the fields $H_s$ required to achieve full
spin polarization of the electrons, calculated using Eq.(3).  The solid line is a linear
fit to the data for $H_{sat}$.
\vspace{0.10in}
}
\label{3}

The solid circles shown in Fig. 3 (b) denote $H_s$ calculated for the data of
Fig. 3 (a), using Eq.(3).  Values of $H_{sat}$ denoted by the open circles are obtained 
from the saturation of the in-plane magnetoconductivity, as illustrate in the inset to Fig.
3 (b).  The line is a fit to the data for $H_{sat}$.  In the narrow range where the two
data sets overlap,
$H_{sat} \approx H_s$ within about $5$\%.

In summary, small angle Shubnikov-de Haas measurements taken at a fixed total magnetic
field of $18$ T as a function of the angle $\theta$ between the magnetic field
direction and the electron plane have allowed separate determination of the
populations of the spin-down and spin-up subbands.  Analysis of the data yields the
field $H_s$ required to achieve complete spin polarization of the electrons.  We find
that $H_s$ is the same as the field $H_{sat}$ which signals the saturation of the
conductivity with an accuracy of $5$\% at $n>3 \times 10^{11}$ cm$^{-2}$.

This work was supported by the US
Department of Energy under Grant No.~DE-FG02-84ER45153.  Partial support was
provided by NSF Grant No.~DMR~98-03440.

\end{multicols}

\begin{references}
\bibitem{assumption}  Note that this step entails the assumption that the $g$-value
is independent of magnetic field.
\end{references}
\end{document}